\documentstyle[11pt]{article}

\font\mybb=msbm10 at 12pt
\def\bb#1{\hbox{\mybb#1}}
\def\Z {\bb{Z}}
\def\R {\bb{R}}
 \def\unit{\hbox to 3.3pt{\hskip1.3pt \vrule height 7pt width .4pt \hskip.7pt
\vrule height 7.85pt width .4pt \kern-2.4pt
\hrulefill \kern-3pt
\raise 4pt\hbox{\char'40}}}

\begin{document}

\pagestyle{empty}
\rightline{UG-11/95}
\rightline{September 1995}
\rightline{hep-th/9509145}

\vspace{2truecm}
\centerline{\bf  Duality Symmetries and the Type II String Effective Action}
\vspace{2truecm}
\centerline{{\bf E.~Bergshoeff}\footnote{
Based on talk given at the Trieste conference on {\it $S$-duality and
Mirror Symmetry}, June 1995.}
}
\vspace{.5truecm}
\centerline{Institute for Theoretical Physics}
\centerline{Nijenborgh 4, 9747 AG Groningen}
\centerline{The Netherlands}

\vspace{3truecm}
\centerline{ABSTRACT}
\vspace{.5truecm}
We discuss the duality symmetries of Type II string effective actions
in nine, ten and eleven dimensions. As a by-product we give a covariant action
underlying the ten--dimensional Type IIB supergravity theory.
We apply duality symmetries to construct dyonic Type II string solutions
in six dimensions and their reformulation as solutions of the
ten--dimensional Type IIB theory in ten dimensions.

\vfill\eject
\pagestyle{plain}

\section{INTRODUCTION}

An important application of duality symmetries in string theory \cite{Gi1}
is
their use as solution-generating transformations at the level
of the low-energy string effective action. To be precise, given a solution
to the (lowest order in $\alpha^\prime$) string equations of motion
with one or more isometries, the duality symmetries generate  new ``dual''
solutions. For this purpose it is important to understand all the
symmetries of the string effective action in the presence of isometries.

Symmetries in the presence of isometries can be understood most easily
by dimensionally reducing the string effection action over the isometry
directions \cite{Mah1}. In general the dimensional reduced action has more
symmetries than the original one. These extra symmetries come
from the following two sources:

\begin{itemize}

\item[1.] General linear transformations over the isometry directions.
In case we dimensionally reduce $n$ directions this leads to a $GL(n,\R)$
symmetry.

\item[2.] Electric-Magnetic duality transformations. These extra symmetries
may occur each time the $D$-dimensional reduced action contains a
$p$-form antisymmetric tensor with $p+1=D/2$. Note that this symmetry
is only realized on the equations of motion.
In the context of string effective actions this symmetry enhancement
happens for dimensions $D \le 8$.

\end{itemize}

The complete group of duality symmetries in the case of Type II theories is
usually called $U$--duality \cite{Hu1}. It contains the target space
duality ($T$--duality) and the strong/weak coupling duality ($S$--duality).

In the first part of this talk I will discuss the different kinds of
Type II duality symmetries and their interrelationships
for the simplest case of one isometry. We are thus naturally led to consider
effective actions in nine and ten dimensions. Since the Type IIA
effective action can be obtained by dimensional reduction of
eleven--dimensional supergravity it is natural to consider eleven dimensions
as well. Further motivations are provided by the fact that eleven-dimensional
supergravity is related to the eleven-dimensional supermembrane \cite{Be1}
and to the ten--dimensional Type IIA superstring
\cite{To3,Wi1}. The discussion below
will be at the classical level and we will be dealing with continous
duality symmetries. It is well-known that these duality symmetries get broken
to discrete subgroups at the quantum level. A discussion
of Type II duality symmetries in $D=9,10,11$ from a more stringy
point of view can be found in \cite{As1}.

\section{DUALITY SYMMETRIES IN $D$= 9,10,11}

We will organize our discussion with increasing number of isometries and
decreasing number of dimensions.  This leads us to consider the following
three cases:

\begin{itemize}

\item[a.] Eleven dimensions without isometries

\item[b.] Eleven dimensions with one isometry. This part
of the discussion is naturally tied up with a consideration of ten dimensions
without isometries.

\item[c.] Eleven dimensions with two isometries. This naturally leads us
to consider the cases of ten dimensions with one isometry and nine
dimensions without isometries as well.

\end{itemize}

Below we will use the results and conventions of \cite{Be2,Be3}.
Note that double hatted fields are eleven--dimensional, hatted fields
are ten--dimensional and unhatted fields are nine--dimensional.
Concerning the $SO^{\uparrow}(1,1)$ scale transformations given below,
it is useful to consider both scale transformations that leave the action
invariant as well as scale transformations that scale the action with
a given weight. The reason for this is that the two types of scale
transformations are related via the process of simple dimensional
reduction. Note that the presence of two scale transformations that
scale the action implies a single scale transformation
that leaves the action invariant.

\subsection{No isometries}

\begin{description}

\item[D=11] There is a single $SO^{\uparrow}(1,1)_{\rm membrane}$ symmetry
that essentially counts the mass dimension of the different fields.
The scale transformations (with continuous parameter $\alpha$)
of the eleven--dimensional fields and action are given by

\begin{equation}
{\hat {\hat g}}^\prime = e^\alpha {\hat {\hat g}}\, ,\ \ \ \ \ \
{\hat {\hat C}}^\prime = e^{3\alpha/2}{\hat {\hat C}}\, ,\ \ \ \ \ \
S^{(11)\prime} = e^{9\alpha/2}S^{(11)}\, .
\end{equation}

We remind
that the three--index gauge field $\hat{\hat{C}}$ is a pseudo-tensor
that changes sign under improper eleven--dimensional g.c.t.'s.

\end{description}

\subsection{One isometry}

We take the isometry to be in the $y$ direction.

\begin{description}

\item[D=11] In addition to the symmetries of the previous subsection, we
have to consider the subgroup of g.c.t.'s that preserve the condition
that the fields do not depend on the coordinate $y$. This group is

\begin{equation}
GL(1,\R)=SO^{\uparrow}(1,1)_y\times\ \Z^{(y)}_{2}\, .
\end{equation}

\noindent Here $SO^{\uparrow}(1,1)_y$ represents the proper coordinate
transformation (with parameter $\beta$) $y \rightarrow e^\beta y$ and
$\Z_2^{(y)}$ represents the improper coordinate transformation
$y \rightarrow -y$.

\item[D=10, Type~IIA] Taking into account that $\hat{\hat{C}}$ changes
sign under the (now) internal $\Z^{(y)}_{2}$, the eleven-dimensional
transformations become the group

\begin{equation}
\label{z2y}
SO^{\uparrow}(1,1)_{\rm membrane} \times\
SO^{\uparrow}(1,1)_{y} \times\ \Z^{(y)}_{2}
\end{equation}

\noindent of global symmetries of the equations of motion. The
scale weights under the different $SO^{\uparrow}(1,1)$'s
and the sign changes under $\Z^{(y)}_{2}$ are given in Table 2,
Appendix B of \cite{Be3}.

\item[D=10, Type~IIB] The underlying Type IIB supergravity theory
will be discussed in more detail in the next section.
Here we only discuss the aspects related
to duality. The Type IIB theory has a $SL(2,\R)_{\rm IIB}$ duality
\cite{Hu1} which in the string frame acts on the fields as follows

\begin{eqnarray}
\label{IIB}
\hat{\jmath}^{\prime}_{\hat{\mu}\hat{\nu}} & = &
|c\hat{\lambda}+d|\ \hat{\jmath}_{\hat{\mu}\hat{\nu}}\, ,\nonumber\\
& &
\nonumber \\
\hat{\lambda}^{\prime} & = & \frac{a\hat{\lambda}+b}{c\hat{\lambda}+d}\, ,\\
& &
\nonumber \\
\left(
\begin{array}{c}
\hat{\cal{B}}^{(1)\prime}_{\hat{\mu}\hat{\nu}} \\
\hat{\cal{B}}^{(2)\prime}_{\hat{\mu}\hat{\nu}} \\
\end{array}
\right)
& = &
\left(
\begin{array}{cc}
d & c \\
b & a \\
\end{array}
\right)
\left(
\begin{array}{c}
\hat{\cal{B}}^{(1)}_{\hat{\mu}\hat{\nu}} \\
\hat{\cal{B}}^{(2)}_{\hat{\mu}\hat{\nu}} \\
\end{array}
\right)\, ,\nonumber
\end{eqnarray}

\noindent where $ad-bc=1$ and $\hat{\lambda}=\hat{\ell}+ie^{-\hat\varphi}$.
We use the convention that $\hat\varphi$ is the
dilaton and $\hat {\cal B}^{(1)}$ the Neveu/Schwarz-Neveu/Schwarz
(NS-NS) axion. Other definitions
of the dilaton and NS-NS axion are possible which differ from
$\hat\varphi$ and $\hat {\cal B}^{(1)}$ by an $SL(2,\R)_{\rm IIB}$ rotation.

There are several interesting subgroups of $SL(2,\R)_{\rm IIB}$.
One is the $(\Z_{4})_{\rm IIB}$ subgroup generated by the element
with $b=1, c=-1$ and $a=d=0$. It leads to the following transformation
rules for the fields:\footnote{The product of two
$\Z_4$ transformations leads to the (trivial) $\Z_2$ transformation
that leaves the coupling constant inert but changes the sign of
both axions.}

\begin{eqnarray}
\label{z4}
\hat{\lambda}^{\prime} & = & -1/\hat{\lambda}\, ,
\hspace{1cm}
\hat{\jmath}^{\prime}_{\hat{\mu}\hat{\nu}}=|\hat{\lambda}|\
\hat{\jmath}_{\hat{\mu}\hat{\nu}}\, ,\nonumber\\
\hat{\cal{B}}^{(1)\prime}_{\hat{\mu}\hat{\nu}} & = &
-\hat{\cal{B}}^{(2)}_{\hat{\mu}\hat{\nu}}\, ,
\hspace{.7cm}
\hat{\cal{B}}^{(2)\prime}_{\hat{\mu}\hat{\nu}}=
\hat{\cal{B}}^{(1)}_{\hat{\mu}\hat{\nu}}\, .
\end{eqnarray}

This transformation inverts the string coupling constant for
$\hat{\ell}=0$, i.e.~$(e^{\hat {\varphi}})^\prime =  1/e^{\hat \varphi}$
when  $\hat \ell = 0$ \cite{Be2}.
It therefore makes sense to identify $SL(2,\R)_{\rm IIB}$
as an $S$-duality group of the Type IIB superstring \cite{Hu1}.
We note that the NS-NS axion $\hat{\cal{B}}^{(1)}$ and the
Ramond-Ramond (R-R) axion $\hat{\cal{B}}^{(2)}$
are rather different in nature. On the one hand, the
NS-NS axion is an elementary excitation whose coupling to the Type IIB
superstring is described by a two-dimensional sigma model.
On the other hand, the R-R axion has its origin in
solitonic modes on the worldsheet and its coupling is not described
by a similar sigma model. Note that under the $(\Z_4)_{\rm IIB}$ transformation
(\ref{z4}) the NS-NS axion is rotated into the R-R axion and {\it vice-versa}.
In view of this the $(\Z_4)_{\rm IIB}$ transformation, besides a ``strong/weak
coupling''
side, also has an ``electric-magnetic'' side from the
worldsheet point of view \cite{Be2}.

We remark that the $(\Z_4)_{\rm IIB}$ strong/weak coupling duality (\ref{z4})
can be converted into a $(\Z_2)_{\rm IIB}$ strong/weak coupling duality
by multiplying it with a so-called $\tilde {\Z}_2^{(y)}$ transformation.
The latter transformation can be obtained from the $\Z_2^{(y)}$ of the
Type IIA superstring given in (\ref{z2y}) by application of the
Type~II Buscher $T$--duality \cite{Dai1,Di1} whose explicit rules are
given in \cite{Be2}\footnote{Note that the Type II $T$
duality rules of \cite{Be2}
are {\it only} valid in the presence of an isometry (in ten dimensions).
Such an isometry has not yet been
assumed in this subsection (it will be done in the next subsection).
Nevertheless the $\tilde {\Z}_2^{(y)}$ rules of the Type IIB theory
can be obtained from the ${\Z}_2^{(y)}$ rules of the Type IIA theory
in this way. In hindsight, the fact that one
obtains the right rules, means that the ${\Z}_2^{(y)}$ and
$\tilde {\Z}_2^{(y)}$  rules, which in themselves do not
require the presence of an isometry, have the special
property that, assuming
there is an isometry, the two symmetries become {\it identical} after
dimensional reduction over the isometry direction.}.
One thus obtains the rules:

\begin{eqnarray}
\label{z2}
\hat{\lambda}^{\prime} & = & 1/\hat{\lambda}^*\, ,
\hspace{.2cm}
\hat{\jmath}^{\prime}_{\hat{\mu}\hat{\nu}}=|\hat\lambda|\
\hat{\jmath}_{\hat{\mu}\hat{\nu}}\, ,
\nonumber\\
\hat{\cal{B}}^{(1)\prime}_{\hat{\mu}\hat{\nu}} & = &
\hat{\cal{B}}^{(2)}_{\hat{\mu}\hat{\nu}}\, ,
\hspace{.5cm}
\hat{\cal{B}}^{(2)\prime}_{\hat{\mu}\hat{\nu}}=
\hat{\cal{B}}^{(1)}_{\hat{\mu}\hat{\nu}}\, ,\\
\hat {D}^\prime_{\hat\mu\hat\nu\hat\rho\hat\sigma} & = & -
\hat {D}_{\hat\mu\hat\nu\hat\rho\hat\sigma}\, ,\nonumber
\end{eqnarray}
which, for $\hat\ell = 0$, again leads to
$(e^{\hat {\varphi}})^\prime =  1/e^{\hat \varphi}$.
We conclude that one can either represent the strong/weak coupling
duality of the Type IIB theory as a $\Z_2$ transformation (like in
(\ref{z2})) but then
it is not part of $SL(2,\R)_{\rm IIB}$ or one represents
it as a particular $SL(2,\R)_{\rm IIB}$ transformation (like in
(\ref{z4})) but then it is not a $\Z_2$ transformation.

Another interesting subgroup of the $S$--duality group
$SL(2,\R)_{\rm IIB}$
is the scaling group $\widetilde{SO}^{\uparrow}(1,1)_{y}$.
Again it can be obtained from the
$SO^{\uparrow}(1,1)_{y}$ of Type IIA, given in (\ref{z2y}),
by application of a Type II Buscher $T$--duality.
Similarly, one can also translate
the $SO^{\uparrow}(1,1)_{\rm membrane}$ of Type IIA to the Type IIB language.
The results are given in Table 3, Appendix B of \cite{Be3}.

In summary, the total global symmetry group of the Type IIB
equations of motion is given by

\begin{equation}
\label{total}
SL(2,\R)_{\rm IIB}\times\
\widetilde{SO}^{\uparrow}(1,1)_{\rm membrane}
\times\ \tilde{\Z}_{2}^{(y)}\, .
\end{equation}

It is intriguing that these symmetries exactly combine into
a $GL(2,\R)_{\rm IIB}$ group. This is the symmetry group
one would expect if the ten-dimensional type IIB theory could
be obtained by dimensional reduction of a (hypothetical)
twelve-dimensional theory with no global symmetries whatsoever.

\end{description}

\subsection{Two isometries}

We take the two isometries to be in the $x$ and $y$ direction.

\begin{description}

\item[D=11] Upon dimensional reduction to nine dimensions, the
general linear transformations in $xy$ space become the group

\begin{eqnarray*}
GL(2,\R) = SL(2,\R) \times\ SO^{\uparrow}(1,1)_{x+y} \times\Z_{2}^{(x)}
\, .\nonumber
\end{eqnarray*}

The $SL(2,\R)$ group
contains the
subgroup $SO^{\uparrow}(1,1)_{x-y}$ corresponding to the
eleven-dimensional g.c.t.~$x\rightarrow e^{\alpha}x\, ,y \rightarrow
e^{-\alpha}y$. The other factor
$SO^{\uparrow}(1,1)_{x+y}$
corresponds to the eleven-dimensional g.c.t.~$x\rightarrow
e^{\beta}x\, ,y \rightarrow e^{\beta}y$.
Finally, $\Z_{2}^{(x)}$ corresponds to the improper
g.c.t.~$x\rightarrow -x$ or any other
$SL(2,\R)$--rotated version of this.

\item[D=10, Type IIA and B] In the presence of an isometry (in ten
dimensions), the Type IIA and Type IIB theories are related by
Type II $T$ duality \cite{Dai1,Di1,Be3}.
There are other global symmetries which are
not covariant from the ten-dimensional point of view. They
become covariant when we rewrite the theories in nine-dimensional
language and so we will discuss them below.

\item[D=9, Type II] In nine dimensions there is a single Type II
theory whose global symmetry group is given by:

\begin{equation}
SL(2,\R)  \times\ SO^{\uparrow}(1,1)_{x+y}
 \times\ SO^{\uparrow}(1,1)_{\rm membrane} \times\ \Z^{(x)}_{2}\, .
\end{equation}

The $SL(2,\R)$ group is a symmetry of the action.  From the Type IIB
point of view it is the manifest $SL(2,\R)_{\rm IIB}$ symmetry of the original
theory. From the point of view of the Type IIA only the
$SO^{\uparrow}(1,1)_y$ can be realized in the absence
of an isometry (in ten dimensions).
The weights of the different nine-dimensional fields are
summarized in Table 4, Appendix B of \cite{Be3}.

We note that in the presence of an isometry in the $x$--direction
the $(\Z_4)_{\rm IIB}$ transformation (\ref{z4}) can be interpreted as the
special linear transformation $x\rightarrow y, y\rightarrow -x$
in $xy$ space. Similarly, in the presence of an
isometry, the $\Z_2$ transformation (\ref{z2})
corresponds to the improper g.c.t.~$x \rightarrow -y,
y\rightarrow -x$.

\end{description}

\section{$D=10$ TYPE IIB SUPERGRAVITY}

In the next section we will illustrate the
application of duality as a solution-generating transformation
through an example. Since this example involves the ten-dimensional
Type IIB superstring, we collect in this section some useful data
concerning the underlying Type IIB supergravity theory.

It is known \cite{Mar1} that the field equations of $D=10$ type IIB
supergravity \cite{Sc1} cannot be derived from a covariant action.
The only equation of motion that
cannot be obtained from an action is that of the four-form gauge field
$\hat{D}$.  This equation of motion states that the field strength
$\hat{F}$ of $\hat{D}$ is self-dual: $\hat{F}={}^{\star}\hat{F}$.  It
follows that if one sets $\hat{F}=0$ everywhere in the equations of
motion, one should be able to obtain the resulting reduced set of
equations from an action, by varying with respect to all fields but
$\hat{D}$.  This was done in Ref.~\cite{Be2}.

One may go one step further and define a non-self-dual  (NSD) theory
underlying the Type IIB theory
by the property that it has the same field content as the original
theory but $\hat{F}$ is not self-dual and, if one imposes self-duality
in the field equations, one recovers the usual Type IIB equations
of motion \cite{Be4}. Since the self-duality condition has
disappeared one can write down an action for this NSD theory.
A useful property of the NSD action is that,
when properly used, it leads to the correct action for the {\it
dimensionally reduced} type IIB supergravity theory.  We thus may avoid
the dimensional reduction of the ten-dimensional type IIB field
equations which is more complicated.  We will make
use of this property in the next section.

The action corresponding to the NSD theory has been given in \cite{Be4}.
In the string frame and with the notation and
conventions of Ref.~\cite{Be2,Be3,Be4}, the NSD action is given by
$(i=1,2)$:

\begin{eqnarray}
\label{eq:stringynonselfdualaction}
\hat{S}_{\rm NSD-IIB}^{{\rm string}}  &=&
{\textstyle\frac{1}{2}} \int d^{10}x
\sqrt{-\hat{\jmath}}
\biggl\{
e^{-2\hat{\varphi}}
\bigl[ -\hat{R} \left( \hat{\jmath} \right)
+4(\partial\hat{\varphi})^{2} -{\textstyle\frac{3}{4}}
\bigl(\hat{\cal H}^{(1)} \bigr)^{2} \bigr]
\\
&&
-{\textstyle\frac{1}{2}} (\partial\hat{\ell})^{2}
 -{\textstyle\frac{3}{4}} \left( \hat{\cal H}^{(2)} -\hat{\ell}
\hat{\cal H}^{(1)} \right)^{2}
-{\textstyle\frac{5}{6}\hat{F}^{2}}-
{\textstyle\frac{1}{96{\sqrt {-\hat\jmath}}}}\epsilon^{ij}\epsilon\hat{D}
\hat{\cal H}^{(i)}\hat{\cal H}^{(j)}
\biggr\}\, .
\nonumber
\end{eqnarray}

\noindent For the sake of completeness we list the definitions of
the field strengths and gauge transformations for the type IIB fields
$\bigl\{\hat{D}_{\hat{\mu}\hat{\nu}\hat{\rho}\hat{\sigma}},\
\hat{\jmath}_{\hat{\mu}\hat{\nu}},\ \hat{\cal
B}^{(i)}_{\hat{\mu}\hat{\nu}},\ \hat{\ell},\ \hat{\varphi}\bigr\}$:

\begin{eqnarray}
\hat{\cal H}^{(i)}
&
=
&
\partial\hat{\cal B}^{(i)}\, ,
\ \ \ \ \
\delta\hat{\cal B}^{(i)}
=
\partial\hat{\Sigma}^{(i)}\, ,
\nonumber\\
\hat{F}
&
=
&
\partial\hat{D} +{\textstyle\frac{3}{4}}
\epsilon^{ij}\hat{\cal B}^{(i)} \partial\hat{\cal B}^{(j)}\, ,
\\
\delta\hat{D}
&
=
&
\partial\hat{\rho} -{\textstyle\frac{3}{4}}
\epsilon^{ij}\partial\hat{\Sigma}^{(i)} \hat{\cal B}^{(j)}\, .\nonumber
\end{eqnarray}

The NSD theory defined by Eq.~(\ref{eq:stringynonselfdualaction}) has
all the symmetries of the type IIB theory, given in the
previous section (see eq.~(\ref{total}))
plus an additional global $\Z_2$ e-m duality of the ${\hat D}$
field that interchanges $\hat{F}$ and ${}^{\star}\hat{F}$.

The $SL(2,\R)_{\rm IIB}$ symmetries have already been given in
eq.~(\ref{IIB}) of the previous section.
To make the $SL(2,\R)_{\rm IIB}$ invariance of the action more
manifest, it is useful to go to
the Einstein
frame because the Einstein metric is inert under them:

\begin{eqnarray}
\label{IIB-E}
S_{\rm NSD-IIB}^{\rm Einstein}
=
{\textstyle\frac{1}{2}}\int d^{10}x\ \sqrt{-\hat{g}}\
&\biggl\{& -\hat{R}
+{\textstyle\frac{1}{4}}
{\rm Tr}\left(\partial_{\mu}\hat{\cal M}
\partial^{\mu}\hat{\cal M}^{-1}\right)
-{\textstyle\frac{3}{4}}
\hat{\cal H}^{(i)} \hat {{\cal M}}_{ij} \hat{\cal H}^{(j)}
\nonumber
\\
& &
-{\textstyle\frac{5}{6}\hat{F}^{2}}-
{\textstyle\frac{1}{96{\sqrt {-\hat{g}}}}}\epsilon^{ij}\epsilon\hat{D}
\hat{\cal H}^{(i)}\hat{\cal H}^{(j)}
\biggr\}\, ,
\label{IIBactionE}
\end{eqnarray}

\noindent where $\hat{g}_{\hat{\mu}\hat{\nu}}
=e^{-\frac{1}{2}\hat{\varphi}} \hat{\jmath}_{\hat{\mu}\hat{\nu}}$ is the
Einstein-frame metric. The matrix $\hat{\cal M}$ is the $2\times 2$ matrix

\begin{equation}
\hat{\cal M}=
\left(
\hat{\cal M}_{ij}
\right)
=
\frac{1}{\Im {\rm m}\lambda}
\left(
\begin{array}{rr}
|\hat{\lambda}|^{2}        & -\Re {\rm e}\hat{\lambda} \\
-\Re {\rm e}\hat{\lambda}  &               1           \\
\end{array}
\right)\, ,
\end{equation}

\noindent where $\hat{\lambda}=\hat{\ell}+ie^{-\hat{\varphi}}$ is a complex
scalar that parametrizes $SL(2,\R)_{\rm IIB}$.

The action (\ref{IIB-E}) is manifestly invariant under the
$SL(2,\R)_{\rm IIB}$ transformations

\begin{equation}
\hat{\cal H}^{\prime}  =  \Lambda\hat{\cal H}\, ,\ \ \ \ \
\hat{\cal M}^{\prime}  =  \left( \Lambda^{-1}\right)^{T}
\hat{\cal M}\Lambda^{-1}\, .
\label{eq:sl2r}
\end{equation}
Here $\hat {\cal H}$ denotes a 2-vector with ${\hat {\cal H}}^{(1)}$ as
the upper component and ${\hat {\cal H}}^{(2)}$ as the lower component.

\noindent If $\Lambda$ is the $SL(2,\R)_{\rm IIB}$ matrix

\begin{equation}
\Lambda=
\left(
\begin{array}{rr}
 d & c \\
b &  a \\
\end{array}
\right)\, ,
\end{equation}

\noindent the transformation Eq.~(\ref{eq:sl2r}) of the matrix $\hat{\cal
M}$ implies the usual transformation of the complex scalar
$\hat{\lambda}$

\begin{equation}
\hat{\lambda}^{\prime}=\frac{a\hat{\lambda}+b}{c\hat{\lambda}+d}\, .
\end{equation}

Finally, we expect that a NSD type IIB theory including fermions can also be
found.  Of course, the full NSD type IIB action cannot be
supersymmetric since the self-duality constraint has no
fermionic partner.  However, the supersymmetry should be recovered in the
field equations of the Type IIB theory after the (super-) self-duality
contraint is
imposed\footnote{We thank M.~Green for a discussion on this point.}.

\section{DYONIC STRINGS IN SIX DIMENSIONS}

As an application of duality as a solution-generating transformation
we will contruct in this section dyonic string solutions in six
dimensions\footnote{A similar construction of dyonic membranes in eight
dimensions has recently been given in \cite{To2}.}.
The reformulation of these dyonic strings as
solutions of the Type IIB theory in ten dimensions will be discussed
in the next section.
We only consider solutions to the source-free equations of motion.
Without sources, we define a $p$-brane solution to be any solution
with $p$ translational spacelike isometries.

It is well-known that reducing $D=10$ Type IIB supergravity over a
4-torus leads to the nonchiral $D=6$ Type IIA supergravity theory
constructed in \cite{Ta1}\footnote{The reduction of $D=10$ Type IIA
supergravity over a 4-torus has recently be considered in
\cite{Se1}.}. This theory has a noncompact $SO(5,5)$
$U$-duality. For our purposes it is enough to consider
a truncated version of the $D=6$
Type IIA theory in which only a subgroup $SO(2,2)$ of the
full $U$-duality group is realized.

To be explicit, reducing the ten-dimensional Type IIB fields
to six dimensions we consider the
following ansatz for the fields ($\mu,\nu,...$ are
six-dimensional spacetime indices and $m,n,...$ are the four internal
directions):

\begin{eqnarray}
\label{ansatzspecial}
\hat{\jmath}_{\mu\nu} & = &\jmath_{\mu\nu}\, , \ \ \ \ \
\hat{\jmath}_{mn}  =  -e^{G+\varphi/2}\delta_{mn}\,,\nonumber\\
\hat{{\cal B}}^{(i)}_{\mu\nu} & = & {\cal B}^{(i)}_{\mu\nu}\, ,
\\
\hat{D}_{\mu\nu\rho\sigma}& = & D_{\mu\nu\rho\sigma}\, , \ \ \
\hat{D}_{mnpq} =  D_{mnpq}\, ,\nonumber\\
\hat{\ell} & = & \ell\, , \ \ \ \ \
\hat{\varphi}  =  \varphi\, . \nonumber
\end{eqnarray}

\noindent All other components are zero. Concerning the $D$ field,
we observe that both $D_{mnpq}$ and $D_{\mu\nu\rho\sigma}$ (after
dualization) lead
in six dimensions to scalars $D=\epsilon^{mnpq}D_{mnpq}$
and $\tilde D$, respectively. Using a suitable normalization
condition for $\tilde D$ turns the self-duality
constraint into $\tilde{D}=D$, which can now be substituted into the
NSD action.  It is therefore enough and consistent to
only collect the $D_{mnpq}$ terms in the dimensional reduction
and to multiply these terms by a factor two.

The resulting reduced action is

\begin{eqnarray}
S&=&{\textstyle\frac{1}{2}}\int d^6 x\sqrt{-\jmath}\biggl [e^{-2\varphi}
  \bigl [-R(\jmath)
  +4(\partial\varphi)^2 \nonumber
- (\partial\bar{G})^2
-{\textstyle\frac{3}{4}}({\cal H}^{(1)})^2
  \bigr ]\nonumber
\\
&& -{\textstyle\frac{1}{2}}e^{2\bar{G}}(\partial\ell)^2
- {\textstyle\frac{3}{4}}e^{2\bar{G}}({\cal H}^{(2)}-\ell{\cal
H}^{(1)})^2
 - {\textstyle\frac{1}{72}}e^{-2\bar{G}}(\partial D)^2
  +{\textstyle\frac{1}{8}}D{\cal H}^T{\cal L}{}^{\ *}{\cal H}\biggr
]\nonumber\, ,
\end{eqnarray}

\noindent where $({}^{*}{\cal H})_{\mu\nu\rho} =
{\textstyle\frac{1}{6\sqrt{-j}}}
\epsilon_{\mu\nu\rho\alpha\beta\gamma}{\cal
H}^{\alpha\beta\gamma}$, $\bar G \equiv G + \varphi/2$,
 and where we have introduced the $2\times 2$
matrix

\begin{equation}
{\cal L}=\left(\matrix{0 & 1 \cr -1 & 0\cr}\right)\, .
\end{equation}

To make the symmetries manifest, it is convenient to go to the
six-dimensional Einstein metric
$g_{\mu\nu}=e^{-\varphi}\jmath_{\mu\nu}$. We thus obtain

\begin{eqnarray}
\label{slraction}
S={\textstyle\frac{1}{2}}\int d^6 x\sqrt{-g}&\biggl[& -R(g)\nonumber
+{2\partial\lambda\partial\bar{\lambda}\over(\lambda-\bar{\lambda})^2}
  +{2\partial\kappa \partial\bar{\kappa }\over(\kappa -\bar{\kappa })^2}
\nonumber\\
&&-\kappa_2{\cal H}^T{\cal M}{\cal H}
  +\kappa_1{\cal H}^T{\cal L}{}^{\ *}{\cal H}\biggr]\, .
\end{eqnarray}

\noindent The complex scalars $\lambda$ and $\kappa$ are defined by

\begin{eqnarray}
\kappa&=&\kappa_1 + i\kappa_2
={\textstyle\frac{1}{8}}D+{\textstyle\frac{3}{4}}ie^{2G}\, , \nonumber
\\ \lambda &=& \lambda_1 + i \lambda_2 = \ell + i e^{-\varphi}\, .
\end{eqnarray}

We see that there are two $SL(2,\R)/U(1)$ scalar cosets in the action
(\ref{slraction}) and correspondingly there are two $SL(2,\R)$
symmetries of the equations of motion.  One of them is the original
$SL(2,\R)_{\rm IIB}$ symmetry of the NSD type IIB action
(\ref{eq:stringynonselfdualaction}).  Note that $G$ and not $\bar{G}$ is
$SL(2,\R)_{\rm IIB}$ invariant.  The second is an electro-magnetic
$SL(2,\R)_{\rm
EM}$ duality that acts on the two-form potentials. The latter is
only a symmetry of the equations of motion.  It acts on $\kappa$ and
${\cal H}$ as follows:

\begin{eqnarray}
\label{EM}
\kappa^{\prime} & = & {p\kappa +q\over r\kappa +s}\,,\nonumber\\
{\cal H}_{\mu\nu\rho}^{\prime} & = & (r\kappa_1 +s){\cal H}_{\mu\nu\rho}
+r\kappa_2 {\cal L}\hat{\cal M}{}^{\ *}{\cal H}_{\mu\nu\rho}\,,
\end{eqnarray}

\noindent with $ps-qr=1$.  A particularly interesting $(\Z_2)_{\rm EM}$
subgroup of $SL(2,\R)_{\rm EM}$ is generated by the element with
$q=-3/4, r= 4/3$ and $p=s=0$. Note that the $SL(2,\R)_{\rm EM}$
transformations are similar in form
to the $S$ duality of the heterotic string compactified to four
dimensions\footnote{See {\it e.g.}~the review of \cite{Se2}.}, with
vector fields replaced by two-form fields and with the axion/dilaton
field replaced by $\kappa$.

In summary, we have recovered a

\begin{equation}
SO(2,2) \equiv SL(2,\R)_{\rm EM} \times SL(2,\R)_{\rm IIB}
\end{equation}

\noindent noncompact symmetry of the six-dimensional
equations of motion.

We are now in a position to construct six-dimensional dyonic
string solutions.  Our starting point is the following six-dimensional
solitonic string solution in the string frame \cite{Du1}:

\begin{eqnarray*}
{\bf 1^{(1)}_{(6){\rm m}}}
\left\{
\begin{array}{rcl}
ds^2 &=& (dx^0)^2-(dx^1)^2 - e^{2\phi} (dx^a)^2\, ,\\
       {\cal H}^{(1)}_{abc} &=& \frac{2}{3}
\epsilon_{abcd}\partial^d\varphi\, ,\\
\varphi &=& \varphi(x^a)\, , \hspace{1cm} \Box e^{2\varphi} = 0\, ,
\label{1m(1)}
\end{array}\right.
\end{eqnarray*}

\noindent where $\Box = \delta^{ab}\partial_a\partial_b$.
The superscript $(1)$ indicates that the charge of the solution
is carried by the NS-NS axion ${\cal B}^{(1)}$.
We have parametrized the six-dimensional space by
$x^\mu = (x^0, x^1, x^a), a \in \{6,7,8,9\}$.

We next apply  to the $1^{(1)}_{(6){\rm m}}$ solution given above the
most general $SL(2,\R)_{\rm IIB} \times SL(2,\R)_{EM}$ transformation,
with parameters $a,b,c,d\ (ad-bc=1)$ (see eq.~(\ref{IIB}))
and $p,q,r,s\ (ps-qr=1)$ (see eq.~(\ref{EM})),
respectively.
The result is given by

\begin{eqnarray}
\label{fam6}
ds^2 &=& A\left[(dx^0)^2 - (dx^1)^2 - e^{2C}(dx^a)^2\right]\, , \nonumber\\
\left(
\begin{array}{c}
{\cal H}^{(1)}\\
       \\
{\cal H}^{(2)}
\end{array}
\right)
&=&
\left(
\begin{array}{c}
dsH -{\textstyle\frac{3}{4}}cre^{-2C}{}^{\ *}H\\
                                   \\
bsH -{\textstyle\frac{3}{4}}are^{-2C}{}^{\ *}H
\end{array}
\right)
\,,\nonumber\\
\ell &=& {bd+ace^{-2C}\over
          d^2 + c^2e^{-2C}}\,,\nonumber\\
e^{-\varphi} &=& {e^{-C}\over d^2+c^2e^{-2C}}\,,\\
D &=& 8{qs
+{\textstyle\frac{9}{16}}pre^{-2C} \over
s^2+{\textstyle\frac{9}{16}}r^2e^{-2C}}\, ,\nonumber\\
e^{2G} &=& {e^{-C}\over s^2 +
{\textstyle\frac{9}{16}}r^2e^{-2C}}\, ,\nonumber
\end{eqnarray}

\noindent where $A$ and $H_{abc}$ are functions of $C$

\begin{eqnarray}
A &=& {\sqrt {d^2+c^2e^{-2C}}}{\sqrt
{s^2 +{\textstyle\frac{9}{16}}r^2e^{-2C}}}\,,\nonumber\\
H_{abc} &=&
{\textstyle\frac{2}{3}}\epsilon_{abcd}\partial^dC\,,
\end{eqnarray}

\noindent and $C$ depends only on the $x^{a}$'s and satisfies
$\Box e^{2C}=0$.

A characteristic feature of the above dyonic string solutions is that
non-zero R-R fields are needed in order for the solution to carry
electric as well as magnetic charge.  Setting the R-R axion ${\cal B}^{(2)}$
and the other
R-R fields $\ell$ and $D$ equal to zero, leads to a purely
electric or magnetic solution.

The family of dyonic string solutions (\ref{fam6}) contains four
purely electrically or magnetically charged string solutions as special
cases\footnote{Strictly speaking there are four more solutions
that can be obtained by acting with a $(\Z_4)^2_{\rm IIB}$ transformation
on the four solutions given below. These extra
solutions only differ in the sign of the axions.}.
First of all, for $a=d=p=s=1,
b=c=q=r=0$ (unit transformation) we recover the original solution
$1^{(1)}_{(6){\rm m}}$.

Secondly, one may perform the $(\Z_4)_{\rm IIB}$ transformation (\ref{z4})
which corresponds to the case $c=p=s=1, b=-1, a=d=q=r=0$.
When acting on the $1^{(1)}_{(6){\rm m}}$ solution,
it leads to the
electrically\footnote{We define the R-R charge of this solution to be
of electric
character because the string coupling constant $e^{\varphi}$ is small
($e^{-2\varphi}$ is singular). According to this definition
{\it all} solutions (both the ones that carry a NS-NS charge and the
ones that carry a R-R charge) have the property that
the dilaton corresponding to the electrically (magnetically)
charged solutions satisfies $\Box e^{-2\varphi} = 0\ (
\Box e^{2\varphi}=0)$. In this way the electric and magnetic solutions
are always connected via a strong/weak coupling duality.}
 charged solution $1_{(6){\rm
e}}^{(2)}$:

\begin{eqnarray*}
{\bf 1^{(2)}_{(6){\rm e}}}\left\{
\begin{array}{rcl}
ds^2 &=& e^\varphi\left[ (dx^0)^2 - (dx^1)^2 \right]
- e^{-{\varphi}}(dx^a)^2\,,\\
{\cal H}^{(2)}_{abc} &=&
{\textstyle\frac{2}{3}}\epsilon_{abcd}\partial^d {\varphi}\, ,
\ \ \ \ \ \ \
e^{2G} = e^{\varphi}\, ,\\
\varphi &=& \varphi(x^a)\, , \hspace{1cm} \Box e^{-2\varphi}=0\, .
\nonumber
\label{fivem2}
\end{array}\right.
\end{eqnarray*}

Thirdly, a $1_{(6)\rm e}^{(1)}$ solution may be obtained
by applying the $(\Z_{4})_{\rm IIB}\times (\Z_{2})_{\rm EM}$
transformation $c=1,b=-1,q=-3/4,r=4/3,a=d=p=s=0$ on the $1_{(6){\rm
m}}^{(1)}$ solution. It is given by \cite{Da1}

\begin{eqnarray*}
{\bf 1^{(1)}_{(6){\rm e}}}
\left\{
\begin{array}{rcl}
ds^2 &=& e^{2\varphi}\left[(dx^0)^2-(dx^1)^2\right] - (dx^a)^2\, ,\\
       {\cal B}^{(1)}_{01} &=& e^{2\varphi}\, ,\\
\varphi &=& \varphi(x^a)\, , \hspace{1cm} \Box e^{-2\varphi} = 0\, .
\label{1e(1)}
\end{array}\right.
\end{eqnarray*}

Finally, by applying the $(\Z_2)_{\rm EM}$ transformation
$a=d=1,q=-3/4,r=4/3, b=c=p=s=0$ on the $1_{(6){\rm m}}^{(1)}$ solution
we obtain a second magnetically charged solution $1_{(6){\rm m}}^{(2)}$:

\begin{eqnarray*}
{\bf 1^{(2)}_{(6){\rm m}}}\left\{
\begin{array}{rcl}
ds^2 &=& e^{-\varphi}\left [ (dx^0)^2 - (dx^1)^2\right ] -
e^{\varphi}(dx^a)^2\,,\\
B_{01}^{(2)} &=& -e^{-2\varphi}\,,
\ \ \ \ \ \ \ \ \ \
e^{2G} = e^{\varphi}\, ,\\
\varphi &=& \varphi(x^a)\, ,\hspace{1cm} \Box
e^{2\varphi}=0\, .
\label{fivee2}
\end{array}\right.
\end{eqnarray*}

It is interesting to compare the six-dimensional dyonic string
solutions (\ref{fam6}) with the
six-dimensional dyonic string solution that was recently constructed in
\cite{Du2}.  Our solution differs from that of \cite{Du2} in the
following two respects.  First of all, the dyonic string of \cite{Du2}
contains no R-R fields whereas our solution does.  For instance, our
generic solution contains two axions while the one of \cite{Du2} contains one.
Secondly, the solution of \cite{Du2} is a heterotic solution that breaks
3/4 of the spacetime supersymmetries.  Our solution is a Type II
solution that breaks 1/2 of the Type II spacetime supersymmetries.  This
is necessarily so because the solitonic string we started with
has this property and we know that the $SL(2,\R)_{\rm
IIB}\times SL(2,\R)_{\rm EM}$ transformation, viewed as a noncompact
symmetry of six-dimensional supergravity, is consistent with the full set
of type II supersymmetries.

\section{STRING/FIVE-BRANES IN $D=10$}

The six-dimensional dyonic string solutions
constructed in the previous section can be reformulated as solutions of the
equations of motion corresponding to the Type IIB theory
in ten dimensions. This can be done in a straightforward manner by using
the relation (\ref{ansatzspecial}) between the six-dimensional and
ten-dimensional fields. The explicit form of the ten-dimensional solutions
can be found in \cite{Be4} and will not be repeated here.

Sofar the discussion has been restricted to the source-free equations
only. The situation with the source terms is less clear. In this
context, see also \cite{To2,Sc2,Ts1}. A complicating feature is that
our solutions are intrinsic Type II solutions, i.e.~with a non-zero
R-R axion. One cannot treat this field as the NS-NS axion
whose coupling to the superstring is described
by a two-dimensional sigma model. For the solutions with
$p=s=1$ and $q=r=0$, an alternative approach is to {\it define}
the NS-NS axion to be the $SL(2,\R)_{\rm IIB}$
rotation of $\hat {{\cal B}}^{(1)}$ such that the NS-NS axion carries all
the charge and the R-R axion none.

Restricting ourselves for the moment to the four special solutions
of the previous section, one may verify
that, including the source terms, they correspond to
the following string and fivebrane solutions of the ten-dimensional
Type IIB theory \cite{Da1,Du5,Ca1,Hu1}:

\begin{eqnarray}
1^{(1)}_{(6)\rm e} &\rightarrow& 1^{(1)}_{(10)\rm e}\, ,
\ \ \ \ \
1^{(1)}_{(6)\rm m} \rightarrow 5^{(1)}_{(10)\rm m}\, ,\nonumber\\
1^{(2)}_{(6)\rm e} &\rightarrow& 5^{(2)}_{(10)\rm e}\, ,
\ \ \ \ \
1^{(2)}_{(6)\rm m} \rightarrow 1^{(2)}_{(10)\rm m}\, .
\end{eqnarray}

\noindent
In relating the ten-dimensional solutions to the six-dimensional
solutions, the ten-dimensional strings are
reduced over four {\it transverse} directions where\-as the five-branes
are reduced over four {\it world-volume} directions.
We observe that some of the six-dimensional string solutions
have a natural interpretation as a ten-dimensional string  whereas
others become five-branes in ten dimensions. It would be interesting
to see whether the ten-dimensional solutions with $a=d=1$ and $b=c=0$
would provide for a natural interpolation between strings and five-branes
in ten dimensions thereby giving further evidence for a
ten-dimensional string/fivebrane duality of the type proposed in
\cite{St1}, like it has been
suggested in \cite{To2} for the case of the conjectured
$D=11$ membrane/fivebrane duality \cite{Hu1,To4}.

Finally, a subset of the ten-dimensional solutions,
corresponding to the case $p=s=1$ and $q=r=0$ has been considered
recently from a more stringy point of view and shown to describe a
whole $SL(2,\Z)$ multiplet of Type IIB superstrings \cite{Sc2}. Note that two
special members of this multiplet are the ten-dimensional
$1^{(1)}_{(10)\rm e}$ and $1^{(2)}_{(10)\rm m}$ solutions
defined above. These two solutions have been used in \cite{Da2,Hu2}
to support evidence for the $D=10$ string/string duality
between the $SO(32)$ heterotic and Type I superstring
proposed in \cite{Wi1}.

\section{DISCUSSION}

The techniques we used in constructing the six-dimensional
dyonic string solutions and their reformulation as ten-dimensional
dyonic string/five-brane solutions
can be applied to more general cases as well.
To explain the basic idea\footnote{The discussion
below is similar to that of \cite{Du3}.}, consider a
supergravity theory containing a metric $g_{\mu\nu}$, a dilaton $\varphi$ and
a $(p+1)$--form gauge field $B_{\mu_1\cdots \mu_{p+1}}$\footnote{
We only consider $p$-brane solutions where the charge is carried
by $B_{\mu_1\cdots \mu_{p+1}}$.  In particular, we do not consider
$p$-brane solutions where the charge is carried by a vector component of
the metric $g$.}.  The Lagrangian for these fields takes a standard
form, as in \cite{Du1}. From the general analysis of \cite{Du1} it
follows that this theory has an elementary $p$-brane solution $p_{\rm
e}$ and a solitonic $(D-p-4)$-brane solution $(D-p-4)_{\rm m}$.  We next
observe that the dual of a $(p+1)$--form field is again a $(p+1)$--form
field in $2(p+2)$ spacetime dimensions.  In order to allow for a $\Z_2$
duality transformation we therefore
reinterpret the $(D-p-4)_{\rm
m}$--solution in $D$ dimensions as a $p_{\rm m}$--solution in $2(p+2)$
dimensions via a dimensional reduction over $D-2p-4$ spacelike {\it
worldvolume} directions.  Similarly, a dimensional reduction of the
$p_{\rm e}$--solution in $D$ dimension over $D-2p-4$ spacelike {\it
transverse} directions leads to a $p_{\rm e}$--solution in $2(p+2)$
dimensions.  Given the standard form of the Lagrangean in $D$
dimensions, and assuming a simple ansatz for the $D$-dimensional fields
that includes the $D$-dimensional $p_{\rm e}$ and $(D-p-4)_{\rm m}$
solutions (as in (\ref{ansatzspecial})),
one can, for all cases we have considered, show that the field equations
corresponding to the dimensionally reduced theory in $2(p+2)$ dimensions
are invariant under a $\Z_2$ duality transformation
that maps the
$p_{\rm e}$ and $p_{\rm m}$ solutions into each other.  In summary, the
special case of the $p_{\rm e}$ solution in $D$ dimensions where there
are $(D-2p-4)$ extra abelian isometries in the transverse directions can
be viewed, via a $\Z_2$ duality transformation
vin $2(p+2)$
dimensions, as the purely ``electrically'' charged partner
of the ``magnetically'' charged $(D-p-4)_{\rm m}$ soliton solution
in $D$ dimensions.

It is instructive to consider a few examples of the above general
analysis.  Consider for instance ten-dimensional type IIA supergravity.
The theory contains a $1,2$-- and $3$--form gauge fields and therefore
has the following solutions (see {\it e.g.}~\cite{Du3} or
the table in \cite{To1}):

\begin{equation}
\bigl( 0_{\rm e}, 6_{\rm m}\bigr)\, ,\hskip .5truecm
\bigl(1_{\rm e}, 5_{\rm m}\bigr)\, , \hskip .5truecm
\bigl(2_{\rm e}, 4_{\rm m}\bigr)\, .
\end{equation}

\noindent Applying the above analysis for $D=10$ and $p=0,1,2$,
respectively, we see that all the elementary solutions can be
reinterpreted as purely ``electrically'' charged parners of the
``magnetically'' charged soliton solutions by a $\Z_2$ duality
transformation in $4,6$ and $8$ dimensions, respectively.

Next, we consider type IIB supergravity.  It contains a complex
$2$--form and a self--dual $4$--form gauge field.  The complex $2$--form
gauge field leads to the following solutions (using the
definition of ``electric'' and ``magnetic'' we adopted for the
R-R axionic charges):

\begin{equation}
\bigl(1_{\rm e} + 1_{\rm m}, 5_{\rm e} + 5_{\rm m}\bigr)\, .
\end{equation}

\noindent In addition, the self--dual $4$--form gauge field leads to the
self--dual threebrane solution $3_{\rm em}$ of \cite{Ho1,Du4}.  By reducing to
six dimensions we find that the $1_{\rm e} + 1_{\rm m}$ solutions
can be considered as the electrically + magnetically
charged partners of the $5_{\rm m} + 5_{\rm e}$ solutions.
This is the case discussed in this talk.

\noindent The self--dual $3_{\rm em}$ solution is unique in the sense
that our general formulae given above lead to a $\Z_2$--duality
transformation in ten dimensions itself.  However, since type IIB
supergravity is already self--dual there is no such
$\Z_2$--transformation.

Finally, we consider the case of eleven-dimensional supergravity.  There
is only one $3$--form in eleven dimensions which leads to an elementary
membrane $2_{\rm e}$ \cite{Du6} and a solitonic five-brane $5_{\rm m}$
\cite{Gu1}. By
performing a $\Z_2$ duality transformation in $8$ dimensions we find
that the $2_{\rm e}$ is the electrically charged partner of the
$5_{\rm m}$ solution.

The results of this work suggest that in all examples given above,
the $\Z_2$ duality
transformation can be extended to an $SL(2,\R)$ transformation in a
relatively simple way.  The $SL(2,\R)$--transformations so obtained can
then be applied to construct dyonic $p$-brane
solutions in $2(p+2)$ dimensions and their reformulation as
$p_e/(D-p-4)_m$ solutions in $D$ dimensions.
In fact, recently a further example has been constructed
for the case of $p=2, D=11$ corresponding to dyonic membranes
in 8 dimensions and their reformulation as membrane/five-brane
solutions in eleven dimensions \cite{To2}.

It would be
interesting to construct further examples of dyonic $p$--brane solutions
and to investigate their properties, like {\it e.g.} their
singularity structure.

\bigskip

\noindent{\bf Acknowledgements}

I would like to thank the organizers of the conference for their
kind invitation to give this talk.
The talk is based on the work described in \cite{Be2,Be3,Be4}
and on the many enlightening discussions I had with my collaborators
Harm Jan Boonstra, Chris Hull, Bert Janssen and Tomas Ort\'\i n.
This work has been made possible by a fellowship of the Royal Netherlands
Academy of Arts and Sciences (KNAW).

\end{document}